\documentclass[aps,plb,twocolumn,showpacs,groupedaddress,floatfix]{revtex4}
\usepackage{graphicx}  
\usepackage{dcolumn}   
\usepackage{bm}        
\usepackage{amssymb}   
\usepackage{multirow}

\begin{document}

\newcommand{\dzero}     {D0}
\newcommand{\ttbar}     {\mbox{$t\bar{t}$}}
\newcommand{\bbbar}     {\mbox{$b\bar{b}$}}
\newcommand{\ccbar}     {\mbox{$c\bar{c}$}}
\newcommand{\herwig}    {{\sc{herwig}}}
\newcommand{\pythia}    {{\sc{pythia}}}
\newcommand{\vecbos}    {{\sc{vecbos}}}
\newcommand{\alpgen}    {{\sc{alpgen}}}
\newcommand{\qq}        {\sc{qq}}
\newcommand{\evtgen}    {{\sc{evtgen}}}
\newcommand{\tauola}    {{\sc{tauola}}}
\newcommand{\geant}     {{\sc{geant}}}
\newcommand{\metcal}    {\mbox{$\not\!\!E_{Tcal}$}}
\newcommand{\met}       {\mbox{$\not\!\!E_T$}}
\newcommand{\pt}        {$p_T$}
\newcommand{\sigmatt}   {\ensuremath{\sigma_{t\bar{t}}}}
\newcommand{\mtop}      {$m_{\rm top}$}
\newcommand{\rsigma}    {\ensuremath{R_{\sigma}}}

\newcommand{\feynhiggs} {{\sc{FeynHiggs}}}
\newcommand{\cpsuperh} {{\sc{CPsuperH}}}

\newcommand{\brh}{\ensuremath{{ B}(t\rightarrow H^+ b)}}

\newcommand{\lumi}      {1~$\rm fb^{-1}$}
\newcommand{\result}    {X.X}
\newcommand{\erstat}    {^{+X.X}_{-X.X}}
\newcommand{\ersyspos}  {+X.X}
\newcommand{\ersysneg}  {-X.X}
\newcommand{\ersys}     {^{\ersyspos}_{\ersysneg}}
\newcommand{\erlumi}    {X.X}
\newcommand{\ljets}     {\ensuremath{\ell}+\rm{jets}}
\newcommand{\eplus}     {\ensuremath{e}+jets}
\newcommand{\muplus}    {\ensuremath{\mu}+jets}
\newcommand{\ltau}      {\ensuremath{\tau \ell}}
\newcommand{\etau}      {\ensuremath{\tau e}}
\newcommand{\mutau}     {\ensuremath{\tau \mu}}

\newcommand{\rs}        {XX.X}
\newcommand{\rserr}     {^{+X.X}_{-X.X}}
\newcommand{\rsigmaell}    {\ensuremath{R_{\ell\ell/\ell j}}}
\newcommand{\rsigmatau}    {\ensuremath{R_{\ltau/\ell \ell\text{-}\ell j }}}


\hspace{5.2in} \mbox{FERMILAB-PUB-09/393-E}
\title{Search for charged Higgs bosons in top quark decays}

\author{V.M.~Abazov$^{37}$}
\author{B.~Abbott$^{75}$}
\author{M.~Abolins$^{65}$}
\author{B.S.~Acharya$^{30}$}
\author{M.~Adams$^{51}$}
\author{T.~Adams$^{49}$}
\author{E.~Aguilo$^{6}$}
\author{M.~Ahsan$^{59}$}
\author{G.D.~Alexeev$^{37}$}
\author{G.~Alkhazov$^{41}$}
% ...

\affiliation{\vspace{0.1 in}(The D\O\ Collaboration)\vspace{0.1 in}}
% ...
\affiliation{$^{6}$University of Alberta, Edmonton, Alberta, Canada;
                Simon Fraser University, Burnaby, British Columbia, Canada;
                York University, Toronto, Ontario, Canada and
                McGill University, Montreal, Quebec, Canada}
% ...
\affiliation{$^{30}$Tata Institute of Fundamental Research, Mumbai, India}
% ...
\affiliation{$^{37}$Joint Institute for Nuclear Research, Dubna, Russia}
% ...
\affiliation{$^{41}$Petersburg Nuclear Physics Institute,
                St. Petersburg, Russia}
% ...
\affiliation{$^{49}$Florida State University, Tallahassee, Florida 32306, USA}
% ...
\affiliation{$^{51}$University of Illinois at Chicago,
                Chicago, Illinois 60607, USA}
% ...
\affiliation{$^{59}$Kansas State University, Manhattan, Kansas 66506, USA}
% ...
\affiliation{$^{65}$Michigan State University,
                East Lansing, Michigan 48824, USA}
% ...
\affiliation{$^{75}$University of Oklahoma, Norman, Oklahoma 73019, USA}
% ...


\date{November 4, 2009}

\begin{abstract}
We present a search for charged Higgs bosons in top quark decays. We analyze the \eplus, \muplus, $ee$, $e\mu$, 
$\mu\mu$, \etau\, and \mutau\, final states from top quark pair production
events, using data from about 1${\text{~fb}}^{-1}$ of integrated
luminosity recorded by the \dzero\ experiment at the Fermilab Tevatron
Collider. We consider different scenarios of possible charged Higgs boson
decays, one where the charged Higgs boson decays purely hadronically
into a charm and a strange quark, another where it decays  into a $\tau$ lepton and a
$\tau$ neutrino  and a third one where both decays
appear. We extract limits on the branching ratio $B(t\rightarrow H^+
b)$ for all these models. We use two methods, one where the $t\bar{t}$
production cross section is fixed, and one where the cross section is
fitted simultaneously with $B(t\rightarrow H^+b)$.  
Based on the extracted limits, we exclude regions in the charged Higgs boson
mass and $\tan \beta$ parameter space for different scenarios of the minimal
supersymmetric standard model.
\end{abstract}

\maketitle 

\section{Introduction} 
In many extensions of the standard model (SM),
including supersymmetry (SUSY) and grand unified theories, the existence of 
an additional Higgs doublet is required. Such models predict multiple physical 
Higgs particles, including three neutral and two charged Higgs bosons
($H^\pm$) \cite{theory_review}. If the charged Higgs boson is sufficiently light, 
it can appear in top quark decays $t\rightarrow H^{+}b$~\cite{noteh}.

Within the SM, the top quark decay into a $W$ boson and a $b$ quark occurs
with almost 100\% probability. The $t\bar{t}$ final state signatures are 
fully determined by the $W$ boson decay modes. 
Measurements of top quark pair production cross sections  
$\sigma_{t\bar{t}}$ in various channels~\cite{xseccombi} 
are potentially sensitive to the decay of top quarks to charged Higgs bosons.
The presence of a light charged Higgs boson would result in 
a different distribution of $t\bar{t}$ events between different  
final states than expected in the SM.  

In this Letter we compare the number of predicted and observed events 
in various $t\bar{t}$ final states and derive 95\% confidence level (CL) 
limits on the production of charged Higgs bosons from top quark decays.   
The analysis is based on data collected with the \dzero\, detector
between August 2002 and February 2006 at the Fermilab Tevatron
{\mbox{$p\bar p$}}\ Collider at {\mbox{$\sqrt{s}$ =\ 1.96\ TeV}}. 
The analyzed datasets correspond to an integrated luminosity of about
$1\text{~fb}^{-1}$. 

The decay modes of the charged Higgs boson depend on the
ratio of the vacuum expectation values of the two Higgs doublets,
$\tan\beta$. For small values of $\tan\beta$ it is dominated by the decay 
to quarks, while for larger values of $\tan\beta$ 
it is dominated by
the decay to a $\tau$ lepton and a neutrino. 
We consider three models for the charged Higgs boson
decay: a purely leptophobic model, where the charged Higgs boson decays into a 
charm and a strange quark, a purely tauonic model, 
where the charged Higgs boson decays exclusively into a $\tau$ lepton
and a neutrino, and a model where both decays can occur. 
In all models we fix the $t\bar{t}$ cross section to the theoretical 
value within the SM and extract \brh. 
In the case of the tauonic model, in addition we
extract $\sigma_{t\bar{t}}$ and \brh\
simultaneously, thus yielding a limit without assuming a particular 
value of the \ttbar~cross section. 

A scenario in which the charged Higgs boson decays exclusively into quarks 
can be realized, for instance, in a general
multi-Higgs-doublet model (MHDM)~\cite{Grossman}. It has been demonstrated
that such leptophobic charged Higgs bosons with a mass of about 80~GeV
could lead to noticeable effects at the Tevatron if $\tan\beta \leq
3.5$~\cite{Akeroyd}. Moreover, large radiative corrections
from SUSY-breaking
effects can lead to a suppression of $H^+\rightarrow\tau^+\nu$
compared to $H^{+}\rightarrow c\bar{s}$~\cite{carena}.
In that case, for small $\tan\beta$, hadronic charged Higgs decays 
can become large in both the two-Higgs-doublet (2HDM)~\cite{Akeroyd}
and the minimal supersymmetric standard model (MSSM).   

For values of  $\tan\beta \ge 20$ we consider different models leading to different branching ratios.
Values of $B(H^{+} \rightarrow c \bar{s})$ close to one are
predicted in specific CP-violating benchmark scenarios (CPX)  
with large threshold corrections~\cite{jae_pilaf}. For other models, 
the tauonic decays of the charged
Higgs boson dominate at high $\tan\beta$, for example, in 
the $m_h^{\rm max}$ benchmark
scenario~\cite{benchmark} where $B(H^+ \rightarrow \tau^+ \nu)$ can be close to one. 

\section{Event selection and analysis method} \label{sec:eventsel}
This search for charged Higgs bosons is based on the following 
$t\bar{t}$ final states:
the dilepton ($\ell\ell$) channel where both
charged bosons ($W^{+}$ or $H^{+}$) decay into a light charged
lepton ($\ell = e$ or $\mu$) 
either directly or through the leptonic decay of a $\tau$, 
the $\tau$+lepton (\ltau) channel where one charged boson decays to a
light charged lepton and 
the other one to a $\tau$-lepton decaying hadronically, and 
the lepton plus jets ($\ell$+jets) channel where one charged boson 
decays to a light charged lepton and the other decays into hadrons. 
We select events to create 14 subchannels:  
({\it i}\,) $ee$ ($\mu \mu$) subchannel with two isolated 
high transverse momentum ($p_T$) electrons (muons) and at least two 
high $p_T$ jets; ({\it ii}\,) $e \mu$ subchannels with one isolated 
high $p_T$ electron and one muon and exactly one or at least two jets; 
({\it iii}\,) \etau\, (\mutau\,) subchannel with one high $p_T$ hadronically 
decaying $\tau$, one electron (muon) and at least two high $p_T$ jets 
one of which is identified as a $b$ jet; ({\it iv}\,) \ljets\ subchannels 
with one isolated high $p_T$ electron (muon), exactly three or at least four   
high $p_T$ jets, further split into subsamples with one or at least  
two $b$-tagged jets.        
Details of the event selection and object reconstruction in the dilepton and \ltau\, 
channels can be found in  Ref.~\cite{Abazov:2009si}; a more detailed 
description of the \ljets\, channel and the combination are given in 
Ref.~\cite{xseccombi}.
All event samples are constructed to be mutually exclusive.  

In the \ljets\ channel the main background consists of $W$+jets
production, with smaller contributions from multijet, single top quark
and diboson production. The background contribution in the \ltau\, channel
is dominated by multijet events, 
while the most important background in the $\ell\ell$ channel emerges
from $Z$+jets production. The sample composition  
of all 14 subchannels, assuming $B(t\rightarrow W^{+}b)=1$ (hence \brh=0), is 
given in Ref.~\cite{xseccombi}.

The simulation of the $W$+jets and $Z$+jets backgrounds as well as the
$t\bar{t}$ signal with no charged Higgs boson decay is performed using
\alpgen~\cite{alpgen} for the matrix element calculation, followed by
\pythia~\cite{pythia} for parton showering and hadronization. Diboson
samples are generated using \pythia, while single top quark events are
simulated using the {\sc{singletop}}~\cite{single_top} generator. 
The generated events are processed through a 
\geant-based ~\cite{geant} simulation of the D0 detector and the same 
reconstruction programs used for the data.

\begin{figure}
\begin{center}
\setlength{\unitlength}{1.0cm}
\begin{picture}(5.0,15.0)
\put(-2.0,0.0){\includegraphics[width=0.42\textwidth,clip=]{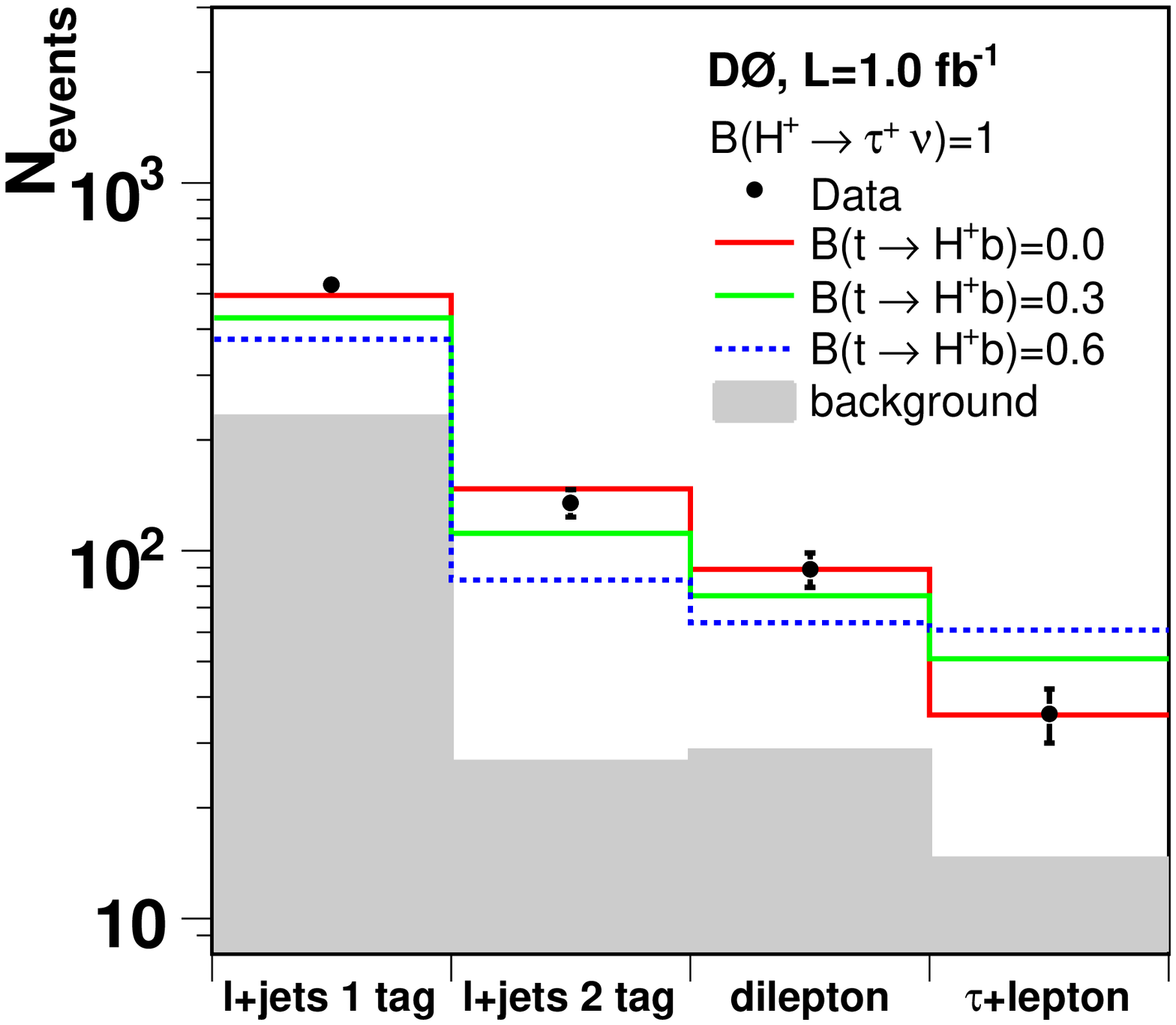}}
\put(-2.0,7.5){\includegraphics[width=0.42\textwidth,clip=]{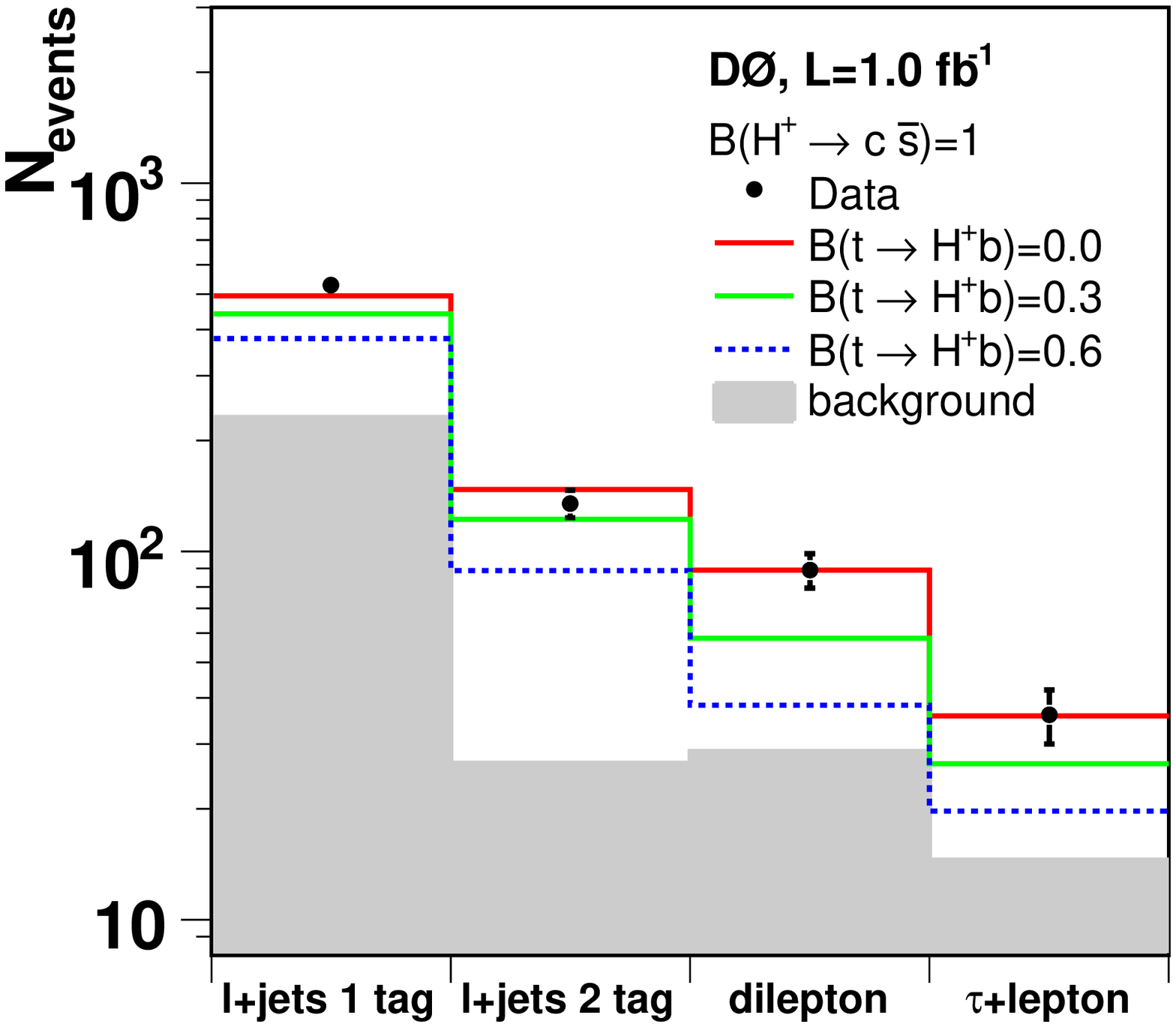}}
\put(0.0,6.34){(b)}
\put(0.0,13.84){(a)}
\end{picture}
\caption{Number of expected and observed events versus final state 
for $M_{H^{+}}=80$~GeV assuming either exclusive $\tau^+\nu$ (a) or
exclusive $c\bar{s}$ (b) decays of the charged Higgs boson.}
\label{fig:illustration_hplus}
\end{center}
\end{figure}

We simulate the signal containing charged Higgs bosons with  
the {\pythia} Monte Carlo event generator~\cite{pythia}, separately for 
the decays $t\bar{t}\rightarrow W^{+}bH^{-}\bar{b}$ (and its charge
conjugate) 
and $t\bar{t}\rightarrow H^{+}bH^{-}\bar{b}$. The total signal 
selection efficiency is calculated as a function of $B\equiv\brh$ as given by:
\begin{eqnarray}
\epsilon_{t\bar{t}} & =&  (1\!-\!B\,)^2\! \cdot\! \epsilon_{t\bar{t}\rightarrow W^{+}bW^{-}\bar{b}}\! +\! 2B\,(1 \!-\! B\,)
\! \cdot\! \epsilon_{t\bar{t}\rightarrow W^{+}bH^{-}\bar{b}} \nonumber \\ 
& & + \, B^2 \! \cdot\!  \epsilon_{t\bar{t}\rightarrow H^{+}bH^{-}\bar{b}} \; ,
\end{eqnarray}
yielding the number of $t\bar{t}$ events as a function of $B$. The 
efficiencies  $\epsilon_{t\bar{t}\rightarrow W^{+}bH^{-}\bar{b}}$ and 
$\epsilon_{t\bar{t}\rightarrow H^{+}bH^{-}\bar{b}}$ are evaluated for 
the assumed $H^{+}$ decay modes.   
Figure~\ref{fig:illustration_hplus} shows the number of  expected
events for different values of \brh\ assuming $M_{H^{+}}=80$~GeV 
and either $B(H^{+} \rightarrow c\bar{s})=1$ or 
$B(H^{+} \rightarrow \tau^{+}\nu)=1$, 
compared to the number of observed events in the considered
channels. The \ljets\ entries with one and two $b$-tags represent 
the sum of four \ljets\ subchannels each (with different light lepton 
flavor and $= 3$ and $\ge 4$ jets). The dilepton contribution corresponds 
to the sum of the $ee$, $\mu\mu$ and two $e\mu$ subchannels, and  
the $\tau$+lepton one shows the sum of the \etau\, and \mutau\, subchannels.     
For a non-zero branching ratio
$B(t\rightarrow H^{+}b \rightarrow c\bar{s} b)$
the number of events decreases in the
\ljets, $\ell\ell$ and \ltau\ final states. 
In case of a non-zero branching ratio
$B(t\rightarrow H^{+}b \rightarrow \tau^+\nu b)$ the number of predicted
events increases in the \ltau\ channel while it decreases in all other
channels. The latter are often called disappearance channels.

\section{\boldmath Extraction of limits on \brh\ } \label{sec:1Dfit}

The extraction of \brh\ is done by calculating the predicted number of events
in 14 search subchannels  
for various charged Higgs boson masses and branching ratios, and  
performing a maximum likelihood fit to the number of observed events in
data. We constrain the multijet background determined from control samples 
in the \ljets\, and \ltau\,
channels by including Poisson terms in the likelihood
function.
We account for systematic uncertainties in the fit by modeling each 
independent source of systematic uncertainty as a Gaussian probability 
density function ${\cal G}$ with zero mean and width corresponding 
to one standard deviation (SD) of the parameter representing the systematic 
uncertainty. 
Correlations of systematic uncertainties between channels
are naturally taken into account by using the same parameter for the
same source of systematic uncertainty. The parameter for each systematic
uncertainty is allowed to float during the likelihood fit.
We maximize the  likelihood function 
\begin{equation} 
{\cal L} = \prod_{i=1}^{14} {\cal P}(n_{i}, m_{i}) 
\times \prod_{j=1}^{14}  {\cal P}(n_{j}, m_{j}) 
\times \prod_{k=1}^{K} {\cal G}(\nu_k;0,{\rm SD}) \,,
\label{eq:mlikeli}
\end{equation}
with ${\cal P}(n,m)$ representing the Poisson
probability to observe $n$ events when $m$ events are expected. The
product runs over the subsamples $i$, and multijet background samples
$j$. $K$ is the total number of independent sources of systematic
uncertainty, with $\nu_k$ being the corresponding nuisance parameter. The
predicted number of events in each channel is the sum of the predicted
background and the expected \ttbar\ events, which depends on \brh. 
   
During the fit, the $t\bar{t}$ cross section is set to
$7.48^{+0.55}_{-0.72}$~pb, corresponding to an approximation to the
next-to-next-to-leading-order (NNLO) QCD cross section that includes
all next-to-next-to-leading logarithms (NNLL) relevant in NNLO
QCD~\cite{moch} at the world average top quark mass of
$173.1$~GeV~\cite{mtop_wa}. 
The uncertainty on the theoretical cross section includes the uncertainty on the 
world average top quark mass.
 
Since we find no evidence for a charged Higgs boson, we extract upper
limits on ${B}(t\rightarrow H^{+}b)$, assuming that $B(H^{+}\rightarrow c\bar{s})=1$,  
or $B(H^{+}\rightarrow \tau^+\nu)=1 $, or a mixture of both. The limit setting procedure follows
the likelihood ratio ordering principle
of Feldman and Cousins~\cite{feldmancousins}. The determination of the
limits requires the generation of pseudo-datasets. We generate ensembles of 
10,000 pseudo-datasets with ${B}(t\rightarrow H^{+}b)$ varied between zero and one  
in steps of 0.05, fully taking into 
account the systematic uncertainties and their correlations.  

Table~\ref{tab:example_fit_80} shows an example of the uncertainties 
on $B(t\rightarrow H^{+}b)$ for $M_{H^{+}}=80$~GeV 
in the tauonic and leptophobic charged Higgs boson models. 
We consider systematic uncertainties originating 
from electron, muon, $\tau$ and jet identification, $\tau$ and 
jet energy calibration, $b$-jet identification, limited statistics 
of data or Monte Carlo samples, modeling of triggers, signal and
background, and integrated luminosity. To evaluate  
the signal modeling uncertainty we replace the SM $t\bar{t}$ sample 
generated with \alpgen\ by the one generated with \pythia\, and 
take the difference in acceptance as systematic uncertainty.  
For both the tauonic and leptophobic model, the two main sources of systematic uncertainty on 
$B(t\rightarrow H^{+}b)$ are the 
uncertainty on the luminosity of 6.1\% and the \ttbar~cross section, followed by the non-negligible uncertainties on signal modeling, $b$~jet identification and jet energy scale. The former two are  approximately of the 
same size as the statistical uncertainty. 
Since in the tauonic model we consider
both appearance and disappearance channels, some
uncertainties affecting the signal and background normalization
cancel. Therefore, uncertainties on signal modeling, the \ttbar~cross
section, lepton identification and luminosity are reduced in the
tauonic model compared to the leptophobic model.

\begin{table}[h]
\begin{center}
\caption{Uncertainties on \brh\ for the  leptophobic and tauonic model, 
assuming  $M_{H^{+}}=80$~GeV.~\label{tab:example_fit_80} }
\begin{tabular}{ccc|cc} \hline
 & \multicolumn{2}{c|}{leptophobic}& \multicolumn{2}{c}{tauonic} \\ \hline
Source                       & +1~SD & $-$1~SD & +1~SD & $-$1~SD\\ \hline
Statistical uncertainty      &     0.057 &    -0.058 & 0.047     &  $-$0.046\\ \hline
Lepton identification        &     0.017 &    -0.017 & 0.010     &  $-$0.010\\
Tau identification           &     0.004 &    -0.004 & 0.006     &  $-$0.006\\
Jet identification           &     0.009 &    -0.009 & 0.010     &  $-$0.010\\
$b$ jet identification       &     0.031 &    -0.030 & 0.030     &  $-$0.030\\
Jet energy scale             &     0.016 &    -0.019 & 0.020     &  $-$0.020\\
Tau energy scale             &     0.004 &    -0.004 & 0.004     &  $-$0.004\\
Trigger modeling             &     0.007 &    -0.011 & 0.007     &  $-$0.006\\
Signal modeling              &     0.023 &    -0.024 & 0.010     &  $-$0.010\\
Background estimation        &     0.013 &    -0.014 & 0.011     &  $-$0.010\\
Multijet background          &     0.014 &    -0.016 & 0.019     &  $-$0.017\\
$\sigma_{t\bar{t}}$          &     0.059 &    -0.085 & 0.040     &  $-$0.054\\
Luminosity                   &     0.056 &    -0.060 & 0.035     &  $-$0.036\\ 
Other                        &     0.017 &    -0.017 & 0.010     &  $-$0.010\\
\hline			    			     			     
Total systematic uncertainty &     0.097 &    -0.118 & 0.071     & $-$0.079 \\
\hline
 \end{tabular}
 \end{center}
\end{table}

\begin{table}[h]
\begin{center}
\caption{Expected and observed upper limits on the branching ratio \brh\ 
for each generated $H^{+}$ mass.~\label{tab:br_limits_fc} }
\begin{tabular}{c|cc|cc|ccc}\hline
&\multicolumn{2}{c|}{leptophobic} & \multicolumn{2}{c|}{tauonic} &
\multicolumn{3}{c}{tauonic from}\\
& & & & &\multicolumn{3}{c}{simultaneous fit}\\
\cline{2-8}
$M_{H^{+}}$ (GeV) & exp & obs  & exp & obs & exp & obs & $\sigma_{t\bar{t}}$~(pb)\\
\hline\\[-8pt]
80   & 0.25 & 0.21 & 0.18 & 0.16 & 0.14 & 0.13 & $8.07^{+1.17}_{-1.04}$\\[2pt]
100  & 0.25 & 0.22 & 0.17 & 0.15 & 0.13 & 0.12 & $8.11^{+1.13}_{-1.00}$\\[2pt]
120  & 0.25 & 0.22 & 0.18 & 0.17 & 0.15 & 0.14 & $8.12^{+1.20}_{-1.05}$\\[2pt]
140  & 0.24 & 0.21 & 0.19 & 0.18 & 0.18 & 0.19 & $8.26^{+1.39}_{-1.20}$\\[2pt]
150  & 0.22 & 0.20 & 0.19 & 0.19 & 0.21 & 0.25 & $8.63^{+1.65}_{-1.38}$\\[2pt]
155  & 0.22 & 0.19 & 0.19 & 0.18 & 0.24 & 0.26 & $8.49^{+1.75}_{-1.45}$\\[1pt]
\hline
\end{tabular}
\end{center}
\end{table}

\begin{figure}
\begin{center}
\setlength{\unitlength}{1.0cm}
\begin{picture}(5.0,11.5)
\put(-2.0,0.0){\includegraphics[width=0.45\textwidth,clip=]{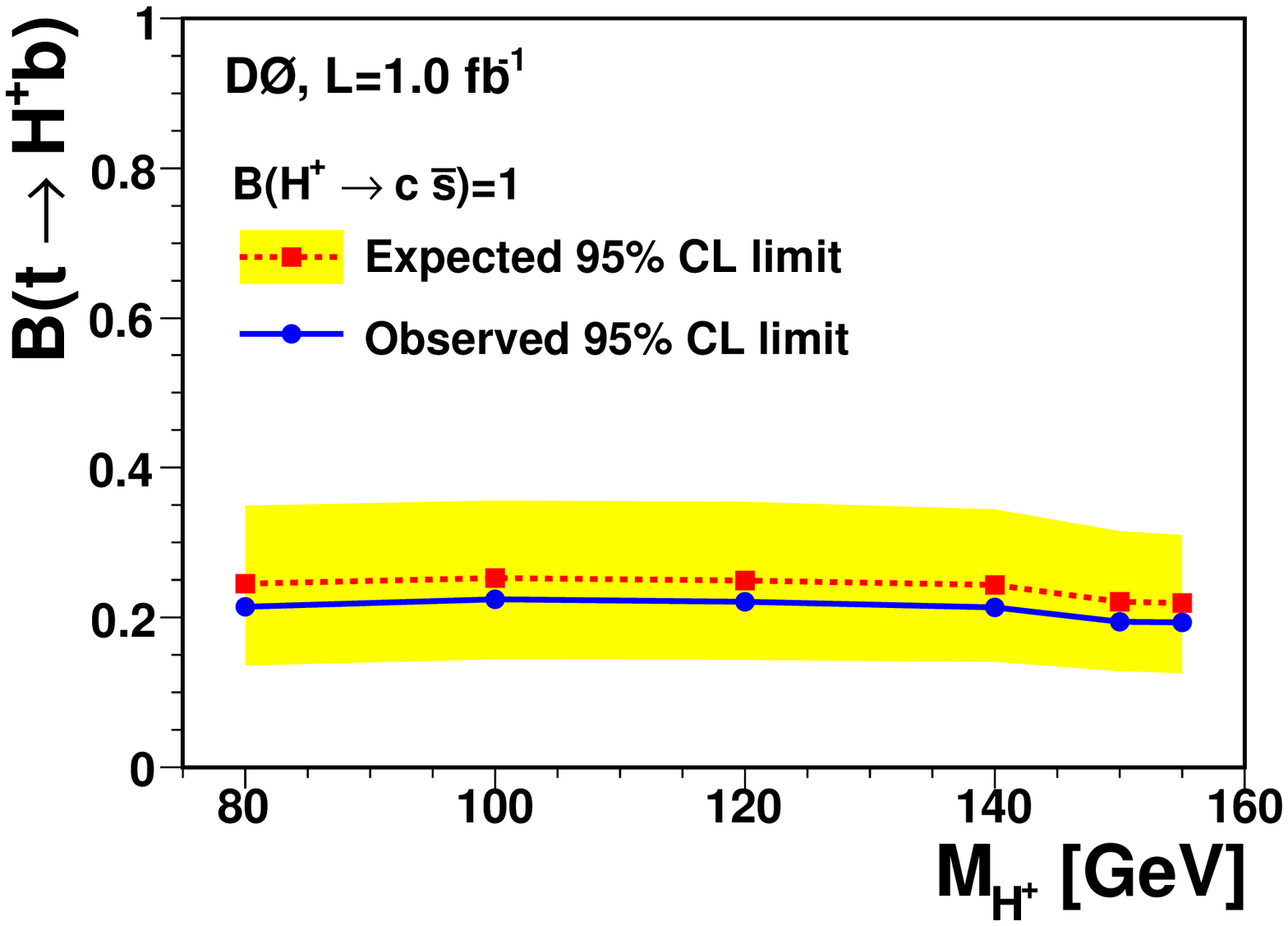}}
\put(-2.0,6.25){\includegraphics[width=0.45\textwidth,clip=]{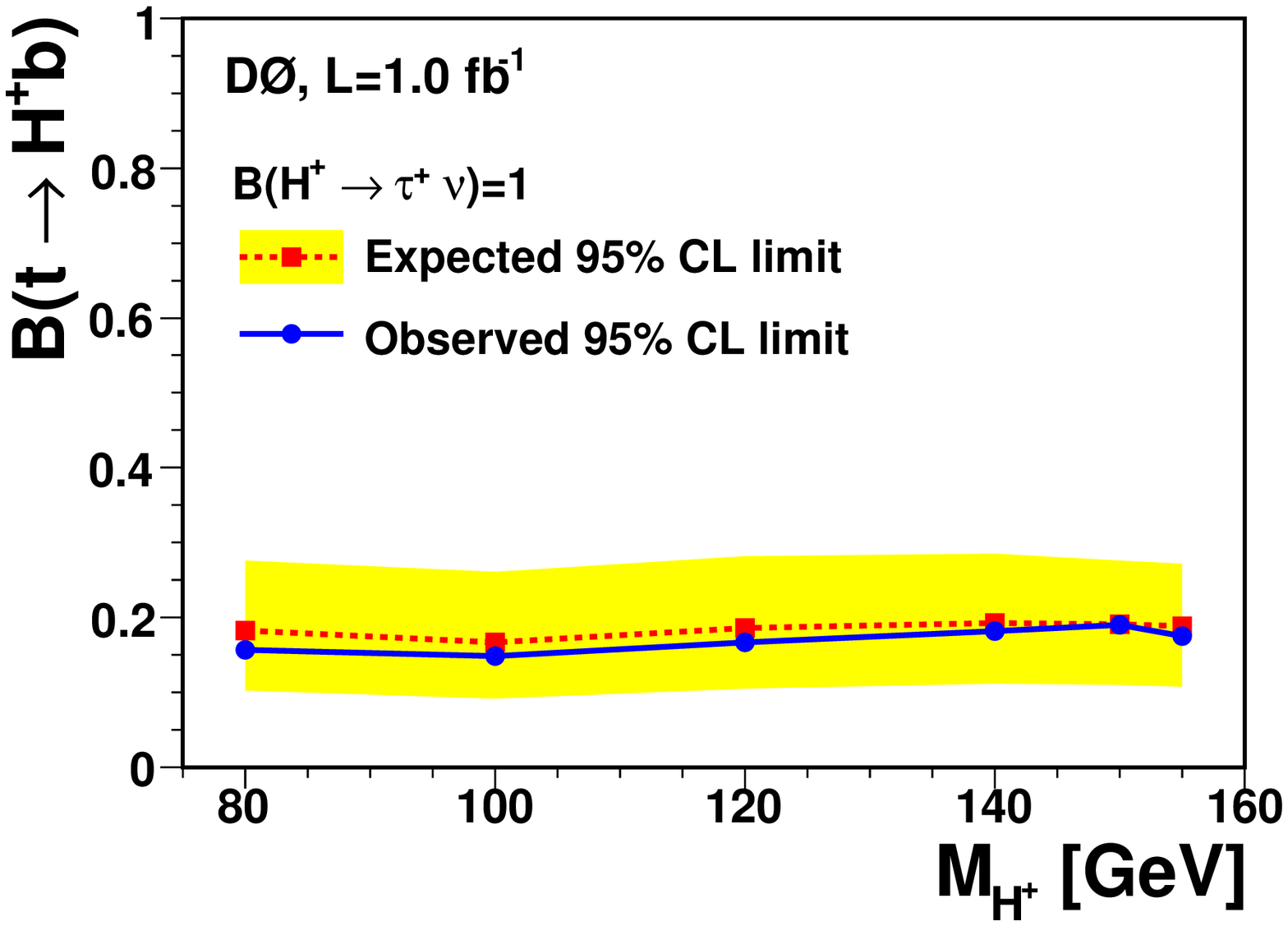}}
\put(4.8,11.02){(a)}
\put(4.8,4.75){(b)}
\end{picture}
\caption{Upper limit on \brh\ for the tauonic (a) and leptophobic (b) model versus $M_{H^{+}}$. 
The yellow band shows the $\pm 1$ SD band around the expected limit (Color version online).}
\label{fig:ljets_dilep_taulep_FC_limits}
\end{center}
\end{figure}

Figure~\ref{fig:ljets_dilep_taulep_FC_limits} 
shows the expected and observed upper limits on 
$B(t\rightarrow H^{+}b)$ assuming 
$B(H^{+} \rightarrow c\bar{s})=1$ or $B(H^{+} \rightarrow \tau^+\nu)=1$ 
as a function of $M_{H^{+}}$  along with the one standard deviation band
around the expected limit. Table~\ref{tab:br_limits_fc} lists the
corresponding expected and observed upper limits on \brh\ for each
generated $M_{H^{+}}$. In the tauonic model we exclude 
$B(t\rightarrow H^{+}b)>0.15-0.19$ and $B(t\rightarrow
H^{+}b)>0.19-0.22$ in the leptophobic case. 

The CDF Collaboration reported a search for charged Higgs bosons using 
different \ttbar\ decay channels  
with a data set of about  $200{\text{~pb}}^{-1}$~\cite{CDF_hp}, resulting in 
$B(t\rightarrow H^{+}b)<0.4$ within the tauonic model. 
Recently, \dzero\, reported limits on \brh\, for the tauonic and
leptophobic models extracted from cross section ratios \cite{xseccombi} and 
for the tauonic model based on a measurement of the \ttbar~cross section 
in $\ell$+jets channel using topological event information~\cite{hplustopo}.  
Exploring the full set of channels as
presented here improves the limits derived in the cross section ratio
method for the leptophobic and for the tauonic model in the 
high $M_{H^{+}}$ region.

\begin{figure*}
\hskip -27pt
\includegraphics[width=1.05\textwidth,clip=]{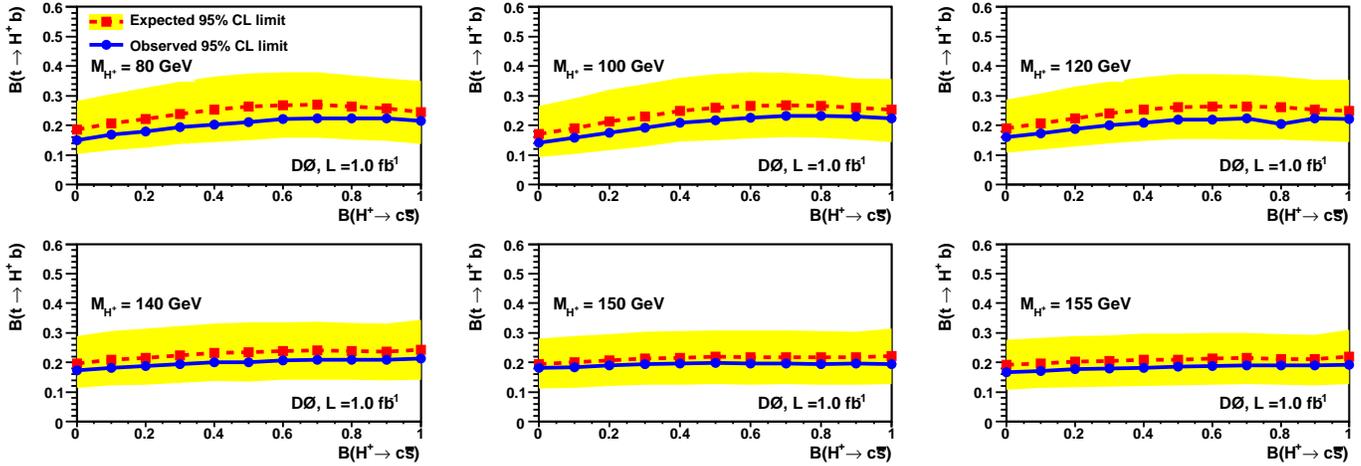}
\caption{Upper limits on $B(t \rightarrow H^{+}b)$ parametrized as function of 
$B(H^{+} \rightarrow c \bar{s})$ for different assumed $M_{H^{+}}$.   
The yellow band shows the $\pm 1$ SD band around the expected limit (Color version online). }
\label{fig:taunu_csbar_limits_mix}
\end{figure*}

We also extract upper limits on $B(t\rightarrow H^{+}b)$
mixing the tauonic and leptophobic models under the
assumption $B(H^{+} \rightarrow \tau^{+} \nu)+B(H^{+} \rightarrow c
\bar{s})=1$. We repeat the extraction of upper limits on $B(t\rightarrow
H^{+}b)$ in the range of $0 \le B(H^{+} \rightarrow \tau^{+} \nu) \le 1$ in
steps of 0.1. For each assumed $M_{H^{+}}$ we parametrize
the expected and observed limits dependent on the mixture between
tauonic and leptophobic decays. 
Figure~\ref{fig:taunu_csbar_limits_mix} shows upper limits on $B(t
\rightarrow H^{+}b)$ as a function of $B(H^{+} \rightarrow c
\bar{s})$. As expected, the upper limit decreases with increasing
tauonic decay fraction.

\section{\boldmath Simultaneous extraction of \brh\ and $\sigma_{t\bar{t}}$} \label{sec:2Dfit}
The search for charged Higgs bosons in top quark decays is based on
the distribution of $t\bar{t}$ events between the various final
states. Naturally, it is also sensitive to the total number of $t\bar{t}$
events. This results in a large systematic uncertainty due to the
theoretical uncertainty in the \ttbar~cross section calculations.
If $\sigma_{t\bar{t}}$ and \brh\ are measured simultaneously the  
limit becomes independent of the assumed theoretical $t\bar{t}$ cross
section. Furthermore, the luminosity uncertainty and other systematic uncertainties affecting the signal normalization are partially absorbed 
by the fitted cross section.

We perform a simultaneous fit of $\sigma_{t\bar{t}}$ and \brh\ for the tauonic 
model. The fitting and limit setting procedure is the same as described in
Sec.~\ref{sec:1Dfit}, with two free parameters instead of one.   
Table~\ref{tab:example_fit_80_2D} shows the uncertainties
on \brh\ and $\sigma_{t\bar{t}}$ for $M_{H^{+}}=80$~GeV. The 
correlation between the two fitted quantities is about 70\% for 
$M_{H^{+}}$ up to 130~GeV and it reaches 90\% for $M_{H^{+}}=155$~GeV. For high 
charged Higgs boson masses, where the correlation becomes high, the sensitivity 
degrades compared to the case where the $t\bar{t}$ cross section is fixed.

The $t\bar{t}$ cross section is set to the measured value in the generation of 
pseudo-datasets for the limit setting procedure. For the fit
to the pseudo-data, $\sigma_{t\bar{t}}$ and \brh\ are allowed to float.
In Table~\ref{tab:br_limits_fc} the expected and observed upper limits
on \brh\ are listed together with the simultaneous measurement of the $t\bar{t}$ cross
section for a top quark mass of 170~GeV. Within uncertainties, the
obtained cross section for all masses of the charged Higgs boson agrees with
$\sigma_{t\bar{t}}=8.18^{+0.98}_{-0.87}$~pb, which was measured on the
same data set assuming $B(t\rightarrow W^{+}b)=1$~\cite{xseccombi}.

\begin{figure}[ht]
\hskip -27pt
\includegraphics[width=0.45\textwidth,clip=]{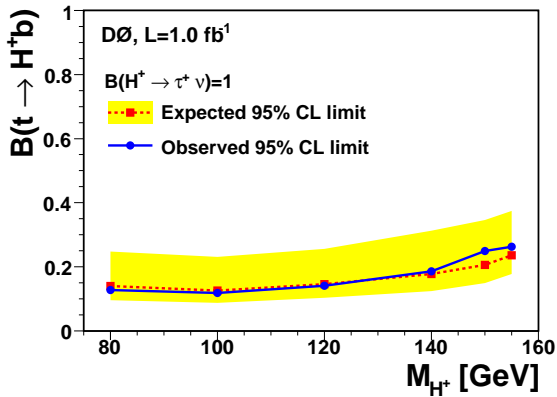}
\caption{Upper limit on \brh\ for the simultaneous fit of \brh\ 
and \sigmatt\, versus $M_{H^{+}}$.
The yellow band shows the $\pm 1$ SD band around the expected limit (Color version online).}
\label{fig:2Dljets_dilep_taulep_FC_limits}
\end{figure}

In Fig.~\ref{fig:2Dljets_dilep_taulep_FC_limits} the upper limits on
\brh~for $M_{H^{+}}$ from $80$ to $155$~GeV are shown. For
small $M_{H^{+}}$, the simultaneous fit provides an
improvement of the sensitivity of more
than 20\% compared to the case where the $t\bar{t}$
cross section is fixed. Furthermore, the  $t\bar{t}$ cross section measured
here represents a measurement independent of the 
assumption $B(t\rightarrow W^{+}b)=1$.

\begin{table}[h]
\begin{center}
\caption{Uncertainties on \brh\ and $\sigma_{t\bar{t}}$ for the simultaneous fit in the tauonic model, 
assuming $M_{H^{+}}=80$~GeV.~\label{tab:example_fit_80_2D} }
\begin{tabular}{ccc|cc} \hline
 & \multicolumn{2}{c|}{ \brh\ }& \multicolumn{2}{c}{$\sigma_{t\bar{t}}$}~(pb) \\ \hline
 Source  &     $+1~SD$    &    $-$1~SD  &     +1~SD    &    $-$1~SD   \\ \hline
                                   Statistical uncertainty & 0.067 &   -0.066 &  0.68 &   -0.64 \\ \hline 
                             Lepton identification & 0.001 & -0.001  &  0.16 &   -0.13 \\ [2pt]
                                Tau identification & 0.014 & -0.014 &  0.12 &   -0.13  \\ [2pt]
                                Jet identification & 0.005 & -0.005  &  0.07 &   -0.07 \\ [2pt]
                                $b$ jet identification & 0.003 & -0.003  &  0.31 &   -0.29\\ [2pt]
                                   Jet energy scale & 0.014 & -0.014 &  0.10 &   -0.09 \\ [2pt]
                                   Tau energy scale & 0.011 & -0.010  &  0.10 &   -0.08  \\ [2pt]
                                           Trigger modeling &  0.009 & -0.000  &  0.12 &   -0.11  \\ [2pt]
                                   Signal modeling & 0.014 & -0.016  &  0.23 &   -0.23 \\ [2pt]
                                    Background estimation & 0.003 & -0.003  &  0.15 &   -0.14 \\ [2pt]
                                 Multijet background & 0.036 & -0.033 &  0.31 &   -0.34 \\ [2pt]
                                       Luminosity &  0.002 & -0.002  &  0.57 &   -0.48 \\  [2pt]
                            Other & 0.006 & -0.006  &  0.17 &   -0.17 \\ [2pt]
                        \hline
                                  Total systematic uncertainty & 0.047 & -0.044  &  0.84 &   -0.77  \\ [2pt]
\hline
 \end{tabular}
 \end{center}
\end{table}

The simultaneous fit requires a reasonably small correlation between 
the two fitted observables. Since at present we have only included disappearance 
channels for the leptophobic model, the correlation between \brh\ and 
$\sigma_{t\bar{t}}$ is large ($\approx$ 90\%) for all charged Higgs boson masses, 
and thus we have not used 
the simultaneous fit method there.

\section{Interpretations in Supersymmetric Models} \label{sec:theory}

The limits on $B(t \rightarrow H^{+}b)$ presented in
Figs.~\ref{fig:ljets_dilep_taulep_FC_limits}, ~\ref{fig:taunu_csbar_limits_mix} 
and ~\ref{fig:2Dljets_dilep_taulep_FC_limits}
are interpreted in different SUSY models and excluded regions of 
$[\tan\beta, M_{H^{+}}]$ parameter space are derived. The investigated
MSSM benchmark models~\cite{benchmark} depend on several model parameters:  
$\mu$ is the strength of the supersymmetric Higgs boson mixing;
$M_{\rm SUSY}$ is a soft SUSY-breaking mass parameter representing a
common mass for all scalar fermions (sfermions) at the electroweak
scale; $A = A_t = A_b$ is a common trilinear
Higgs-squark coupling at the electroweak scale; 
$X_t = A - \mu \cot\beta$ is the  stop mixing 
parameter; $M_2$ denotes a common SU(2) gaugino mass at the
electroweak scale; and $M_3$ is the gluino mass. The top quark mass,
which has a significant impact on the calculations through radiative
corrections, is set to the current world average of
$173.1$~GeV~\cite{mtop_wa}. 

Direct searches for charged Higgs bosons have been performed by the LEP 
experiments resulting into limits of $M_{H^{+}} <79.3$~GeV in the 
framework of 2HDM \cite{LEP_direct}.  
Indirect bounds on $M_{H^{+}}$ in the region of $\tan\beta < 40$ were obtained  
for several MSSM scenarios \cite{LEP_indirect}, two of which are identical to the ones 
presented in Sect.\ref{sec:nomix} and \ref{sec:mhmax} of this Letter.

\begin{table}[h]
\begin{center}
\caption{Summary of the most important SUSY parameter values (in GeV) 
for different MSSM benchmark scenarios.~\label{tab:mssm_para} }
\vspace {0.1 cm}
\begin{tabular}{l|c|c|c}
\hline\\[-9pt]
parameter      & $\rm CPX_{gh}$     & $m_h$-max     & no-mixing \\
\hline
$\mu$           & 2000                     & 200       & 200      \\
$M_{\rm SUSY}$  & 500                      & 1000      & 2000     \\
$A$             & $1000\cdot\exp(i \pi/2)$ &           &          \\
$X_t$           &                          & 2000      & 0        \\
$M_2$           & 200                      & 200       & 200      \\
$M_3$           & $1000\cdot\exp(i \pi)$   & 800       & 1600     \\
\hline
\end{tabular}
\end{center}
\end{table}

\subsection{Leptophobic model}
\begin{figure}[hb]
\centering
\includegraphics[width=0.45\textwidth,clip=]{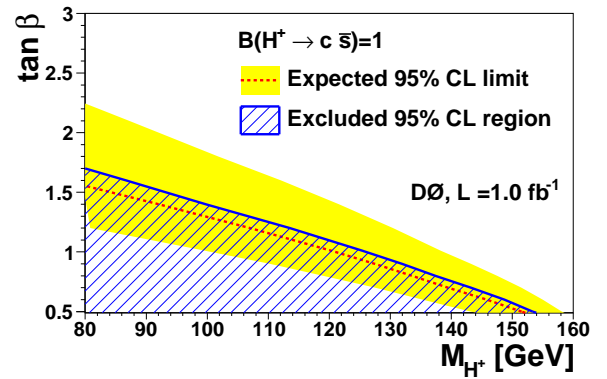}
\caption{Excluded regions of $[\tan\beta,M_{H^+}]$ parameter space for 
leptophobic model. The yellow band shows the $\pm 1$ SD 
band around the expected limit (Color version online).}
\label{fig:mssm_csonly}
\end{figure}

A leptophobic model with a branching ratio of $B (H^{+} \rightarrow
c\bar{s}) = 1$ is possible in MHDM~\cite{Grossman,Akeroyd}. Here we 
calculate the branching
ratio $B(t \rightarrow H^{+}b)$ as a function of $\tan \beta$, and the
charged Higgs boson mass including higher order QCD corrections~\cite{param} 
using \feynhiggs~\cite{feynhiggs}. Figure~\ref{fig:mssm_csonly} shows the
excluded region of $[\tan\beta,M_{H^+}]$ parameter space. For $\tan\beta=0.5$,
for example, $M_{H^{+}}$ up to 153~GeV are excluded.
For low $M_{H^{+}}$, values of $\tan\beta$ up to 1.7 are 
excluded. These 
are the most stringent limits on the $[\tan \beta, M_{H^+}]$ plane in leptophobic
charged Higgs boson models to date. 

\subsection{CPX model with generation hierarchy}
\begin{figure}
\centering
\includegraphics[width=0.48\textwidth,clip=]{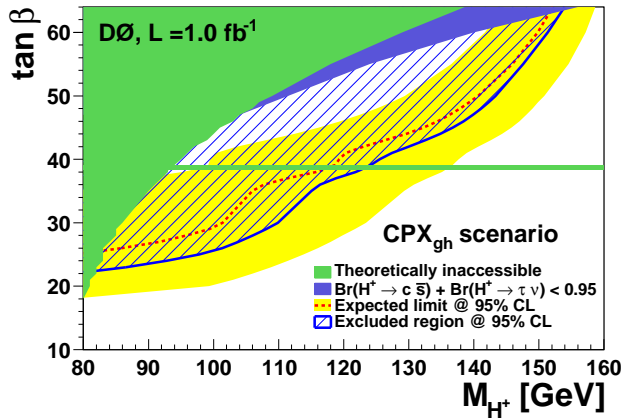}
\caption{Excluded region of $[\tan\beta,M_{H^+}]$ parameter space in the MSSM for
the $\rm CPX_{gh}$ scenario with generation hierarchy such that $B(H^{+}
\rightarrow c \bar{s}) + B(H^{+} \rightarrow \tau^{+} \nu ) \approx 1$.
The yellow band shows the $\pm 1$ SD band around the expected limit (Color version online).}
\label{fig:mssm_cpx}
\end{figure}

$B(H^{+} \rightarrow \tau^{+} \nu) +B(H^{+}
\rightarrow c \bar{s}) \approx 1$ can be realized in a particular
CPX benchmark scenario ($\rm CPX_{gh}$)~\cite{jae_pilaf} of the MSSM. 
This scenario is identical to the CPX scenario investigated in~\cite{LEP_indirect}
except for a different choice of arg($A$) and an additional mass hierarchy 
between the first two and the third generation of sfermions 
which is introduced as follows:
\begin{equation}
M_{\tilde{X}_{1,2}} = \rho_{\tilde{X}} M_{\tilde{X}_{3}}\,,
\end{equation}
where $\tilde{X}$ collectively represents the chiral multiplet for the left-handed doublet
squarks $\tilde{Q}$, the right-handed up-type (down-type) 
squarks $\tilde{U}$ ($\tilde{D}$), the left-handed doublet
sleptons $\tilde{L}$ or the right-handed charged sleptons $\tilde{E}$.
Taking $\rho_{\tilde{U},\tilde{L},\tilde{E}}=1$,
$\rho_{\tilde{Q},\tilde{D}} = 0.4$ and requiring that the masses of the
scalar left- and right-handed quarks and leptons are large
$M_{\tilde{Q}_3,\tilde{D}_3} = 2 M_{\tilde{U}_3,\tilde{L}_3,\tilde{E}_3} = 2$~TeV,  
we calculate the branching ratios $B(t \rightarrow H^{+}b)$ including higher order QCD
and higher order MSSM
corrections using the $\rm CPX_{gh}$ MSSM parameters in Table~\ref{tab:mssm_para}. The
calculation is performed with the program
\cpsuperh~\cite{cpsuperh}. Figure~\ref{fig:mssm_cpx} shows the excluded 
region in the $[\tan\beta,M_{H^{+}}]$ parameter space. Theoretically 
inaccessible regions indicate parts of the parameter space where 
perturbative calculations cannot be performed reliably.  
In the $[\tan\beta,M_{H^+}]$ region 
analyzed here, the sum of the branching ratios was found to be
$B(H^{+} \rightarrow \tau^{+} \nu) +B(H^{+} 
\rightarrow c \bar{s})>0.99$ except for values very close to the blue
region which indicates $B(H^{+} \rightarrow \tau^{+} \nu) +B(H^{+}
\rightarrow c \bar{s})<0.95$. The charged Higgs decay 
$H^{+} \rightarrow \tau^{+} \nu$ dominates for $\tan\beta$ below 22 
and above 55. For the rest of the $[\tan\beta,M_{H^{+}}]$ parameter 
space both the hadronic and the tauonic decays of
charged Higgs bosons are important. In the region $38 \le \tan\beta \le 40$,
the hadronic decays of the charged Higgs boson dominate and 
$ B(H^{+} \rightarrow c \bar{s}) > 0.95$. 
For large values of $\tan\beta$, $M_{H^{+}}$ up to 154~GeV 
are excluded. For low charged Higgs masses, $\tan\beta$ values down to 23 are 
excluded. These are the first Tevatron limits on
a CP-violating MSSM scenario derived from the charged Higgs 
sector.

\subsection{No-mixing scenario \label{sec:nomix}}

\begin{figure}
\centering

\includegraphics[width=0.48\textwidth,clip=]{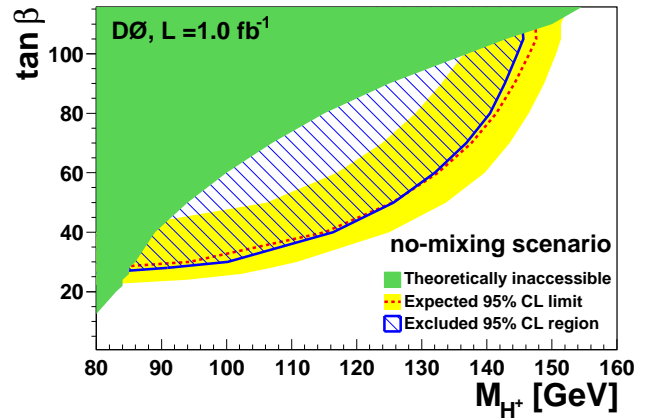}
\caption{Excluded region of $[\tan\beta,M_{H^+}]$ parameter space 
in the MSSM for the 
no-mixing scenario. The yellow band shows the $\pm 1$ SD band 
around the expected limit (Color version online).}
\label{fig:nomixing_tauonly}
\end{figure}

In the CP-conserving no-mixing scenario, the stop mixing
parameter $X_t$ is set to zero, giving rise to a relatively restricted
MSSM parameter space. In the $[\tan\beta,M_{H^+}]$ parameter space 
analyzed here the branching ratio is $B(H^{+} \rightarrow \tau^{+} \nu
)>0.99$ except for very low values of $\tan\beta$ and $M_{H^+}$ where
$B(H^{+} \rightarrow \tau^{+} \nu )>0.95$. We interpret the results
derived in the tauonic model using the simultaneous fit in the framework of the no-mixing
scenario. The branching ratios $B(t\rightarrow H^{+}b)$ are calculated
including higher order QCD and higher order MSSM
corrections using the no-mixing MSSM parameters as given in
Table~\ref{tab:mssm_para}. The calculation is performed with \feynhiggs~\cite{feynhiggs}. 
Figure~\ref{fig:nomixing_tauonly}
presents the excluded region of $[\tan\beta,M_{H^+}]$ parameter space. 
For large values of $\tan\beta$, $M_{H^{+}}$ up to
145~GeV are excluded. 
For low $M_{H^{+}}$, values of $\tan\beta$ down to 27 are 
excluded.

\subsection{$m_h$-max scenario \label{sec:mhmax}}
In the CP-conserving $m_h$-max scenario the stop mixing parameter 
is set to a large value, $X_t=2M_{\rm SUSY}$. 
The theoretical upper bound on the lighter CP-even neutral
scalar, $m_h$, for a given value of $\tan\beta$ and 
fixed $m_t$ and $M_{\rm SUSY}$ is designed to be maximal. Therefore the 
model provides the largest parameter space in $m_h$ and 
as a consequence, less restrictive exclusion limits on $\tan\beta$ than 
the other models. 
In the investigated $[\tan\beta,M_{H^+}]$ parameter space, $B(H^{+}
\rightarrow \tau^{+} \nu)>0.99$ holds except for low values of $\tan\beta$
and $M_{H^+}$, where 
$B(H^{+} \rightarrow \tau^{+} \nu )>0.97$. Thus we use the simultaneous 
fit results within the tauonic model to derive constraints on the $m_h$-max
scenario. The branching ratios $B(t\rightarrow H^{+}b)$ are
calculated using \feynhiggs~\cite{feynhiggs}
including higher order QCD and higher order MSSM
corrections. The $m_h$-max MSSM parameters are given in
Table~\ref{tab:mssm_para}.

\begin{figure}[hb]
\centering
\includegraphics[width=0.48\textwidth,clip=]{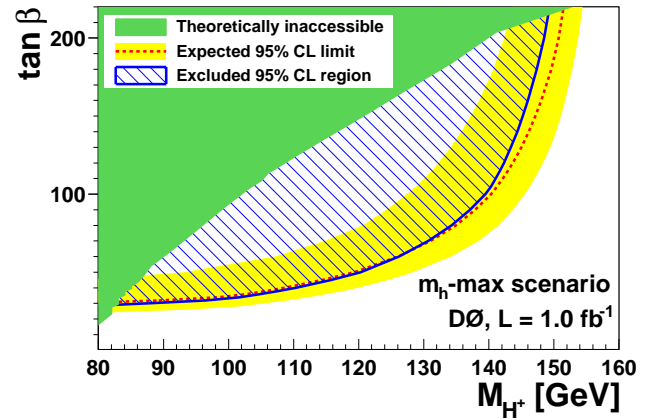}
\caption{Excluded region of $[\tan\beta,M_{H^+}]$ parameter space in the MSSM for 
the $m_h$-max scenario. The yellow band shows the $\pm 1$ SD band 
around the expected limit (Color version online).}
\label{fig:mhmax_tauonly}
\end{figure}

Figure~\ref{fig:mhmax_tauonly} shows the excluded region of 
$[\tan\beta,M_{H^+}]$ parameter space. For large values of 
$\tan\beta$, $M_{H^{+}}$ up to 149~GeV are excluded.
These are the most stringent limits from the Tevatron to date. 
For low charged
Higgs boson masses, values of $\tan\beta$ down to 29 are  
excluded.

\section{summary}
We have performed a search for charged Higgs bosons in top quark decays. No
indication for charged Higgs boson production in the tauonic or leptophobic 
model is found. Upper limits at 95\% CL on the \brh\ branching ratios are 
derived in different scenarios depending on the values of 
$B (H^{+} \rightarrow c\bar{s})$ and $B(H^{+} \rightarrow \tau^{+} \nu )$. 
For the leptophobic model, $B(t\rightarrow H^{+}b)>0.22$ is excluded for the 
$M_{H^{+}}$ range between 80
and 155 GeV. For the tauonic model, $B(t\rightarrow H^{+}b)>0.15-0.19$ are excluded 
depending on $M_{H^{+}}$. 
In this model we have also performed a model-independent measurement
and excluded $B(t\rightarrow H^{+}b)>0.12-0.26$ depending on $M_{H^{+}}$. 

We interpret the results in different models and exclude regions in
$[\tan\beta,M_{H^{+}}]$ parameter space. For the $m_h$-max scenario,
for example, $M_{H^{+}}$ values up to 149~GeV are excluded. 
These are the most restrictive limits to date in direct searches for charged
Higgs boson production in top quark decays.


We thank the staffs at Fermilab and collaborating institutions, 
and acknowledge support from the 
DOE and NSF (USA);
CEA and CNRS/IN2P3 (France);
FASI, Rosatom and RFBR (Russia);
CNPq, FAPERJ, FAPESP and FUNDUNESP (Brazil);
DAE and DST (India);
Colciencias (Colombia);
CONACyT (Mexico);
KRF and KOSEF (Korea);
CONICET and UBACyT (Argentina);
FOM (The Netherlands);
STFC and the Royal Society (United Kingdom);
MSMT and GACR (Czech Republic);
CRC Program, CFI, NSERC and WestGrid Project (Canada);
BMBF and DFG (Germany);
SFI (Ireland);
The Swedish Research Council (Sweden);
CAS and CNSF (China);
and the
Alexander von Humboldt Foundation (Germany).
We would also like to thank J.~S.~Lee and A.~Pilaftsis for providing us with the 
$\rm CPX_{gh}$ model and many stimulating discussions.   



\begin{thebibliography}{99}


\bibitem[a]{alton}
Visitor from Augustana College, Sioux Falls, SD, USA.
\bibitem[b]{askew,atramentov,gershtein}
Visitor from Rutgers University, Piscataway, NJ, USA.
\bibitem[c]{burdin}
Visitor from The University of Liverpool, Liverpool, UK.
\bibitem[d]{luna-garcia}
Visitor from Centro de Investigacion en Computacion - IPN,
  Mexico City, Mexico.
\bibitem[e]{podesta-lerma}
Visitor from ECFM, Universidad Autonoma de Sinaloa, Culiac\'an, Mexico.
\bibitem[f]{weber}
Visitor from Universit{\"a}t Bern, Bern, Switzerland.
\bibitem[g]{wenger}
Visitor from Universit{\"a}t Z{\"u}rich, Z{\"u}rich, Switzerland.

%
\vskip 0.25cm



%


\bibitem{theory_review} D.~J.~H.~Chung, L.~L.~Everett, G.~L.~Kane, S.~F.~King, J.~Lykken  
and L-T. ~Wang, Phys.\ Rept. {\bf 407}, 1 (2005).  

\bibitem{noteh}
Throughout the Letter, $H^{+}$ and $W^{+}$ also refer to the charge conjugate state.


\bibitem{xseccombi}
 V.~M.~Abazov {\it et al.}  [D0 Collaboration],
Phys. Rev. D {\bf80} (2009).



\bibitem{Grossman}
  Y.~Grossman, Nucl.\ Phys.\ {\bf B426}, 355 (1994). 

\bibitem{Akeroyd}
  A.~G.~Akeroyd, arXiv:hep-ph/9509203 (1995).

\bibitem{carena}
 M.~S.~Carena, S.~Mrenna and C.~E.~M.~Wagner,
 Phys.\ Rev.\  D {\bf 62}, 055008 (2000).


\bibitem{jae_pilaf} 
  J.~S.~Lee, Y.~Peters, A.~Pilaftsis and C.~Schwanenberger,
  arXiv:0909.1749 [hep-ph].

\bibitem{benchmark} 
  M.~S.~Carena, S.~Heinemeyer, C.~E.~M.~Wagner and G.~Weiglein,
  Eur.\ Phys.\ J.\  C {\bf 26}, 601 (2003).

\bibitem{Abazov:2009si}
  V.~M.~Abazov {\it et al.}  [D0 Collaboration],
  Phys.\ Lett.\  B {\bf 679}, 177 (2009).

\bibitem{alpgen}
M.L.~Mangano {\sl et al.}, JHEP\ {\bf 07}, 001 (2003).

\bibitem{pythia} T. Sj\"ostrand {\sl et al.}, Comput.
  Phys. Commun. \textbf{135}, 238 (2001).

\bibitem{single_top}
E.E.~Boos {\sl et al.}, Phys.\ Atom.\ Nucl.\ {\bf 69}, 1317 (2006).


\bibitem{geant}
R. ~Brun and F. ~Carminati, CERN Program Library Long Writeup W5013, 1993 (unpublished).

\bibitem{moch} S.~Moch and P.~Uwer,
  Phys.\ Rev.\  D {\bf 78}, 034003 (2008); S.~Moch and P.~Uwer, private communications.

\bibitem{mtop_wa}
  Tevatron Electroweak Working Group and CDF Collaboration and D0
  Collaboration,
  arXiv:0903.2503 [hep-ex] (2009).


\bibitem{feldmancousins}
 G. ~Feldman and R. ~Cousins, Phys. Rev. D \textbf{57}, 3873 (1998).

\bibitem{CDF_hp}
  CDF Collaboration, A.~Abulencia {\sl et al.}, 
  Phys.\ Rev.\ Lett.\  {\bf 96}, 042003 (2006). 

\bibitem{hplustopo}
 V.~M.~Abazov {\it et al.}  [D0 Collaboration],
  Phys.\ Rev.\  D {\bf 80}, 051107 (2009).

\bibitem{LEP_direct}
  OPAL Collaboration, G.~Abbiendi {\sl et al.}, arXiv:0812.0267 [hep-ex]; 
  ALEPH Collaboration, A.~Heister {\sl et al.}, Phys.\ Lett.\  B {\bf 543}, 1 (2002); 
  L3 Collaboration, P.~Achard {\sl et al.}, Phys.\ Lett.\  B {\bf 575}, 208 (2003); 
  DELPHI Collaboration, J.~Abdallah {\sl et al.}, Eur.\ Phys.\ J.\  C {\bf 34}, 399 (2004); 
  ALEPH, DELPHI, L3 and OPAL Collaborations, the LEP working group for Higgs boson searches, 
  arXiv:hep-ex/0107031.

\bibitem{LEP_indirect}
   ALEPH, DELPHI, L3 and OPAL Collaborations, S.~Schael {\sl et al.}, 
  Eur.\ Phys.\ J.\  C {\bf 47}, 547 (2006).
\bibitem{param} 
Since this model cannot be realized in the MSSM without further modifications,
higher order SUSY corrections are not included.

\bibitem{feynhiggs}
  M.~Frank, T.~Hahn, S.~Heinemeyer, W.~Hollik, H.~Rzehak and G.~Weiglein,
  JHEP {\bf 0702}, 047 (2007);
  G.~Degrassi, S.~Heinemeyer, W.~Hollik, P.~Slavich and G.~Weiglein,
   Eur.\ Phys.\ J.\  C {\bf 28}, 133 (2003);
  S.~Heinemeyer, W.~Hollik and G.~Weiglein,
  Eur.\ Phys.\ J.\  C {\bf 9}, 343 (1999);
  S.~Heinemeyer, W.~Hollik and G.~Weiglein,
  Comput.\ Phys.\ Commun.\  {\bf 124}, 76 (2000).

\bibitem{cpsuperh}
 J.~S.~Lee, M.~Carena, J.~Ellis, A.~Pilaftsis and C.~E.~M.~Wagner,
 Comput.\ Phys.\ Commun.\  {\bf 180}, 312 (2009); 
M.~S.~Carena, J.~R.~Ellis, A.~Pilaftsis and C.~E.~M.~Wagner,
Nucl.\ Phys.\ {\bf B586}, 92 (2000); 
M.~S.~Carena, J.~R.~Ellis, A.~Pilaftsis and C.~E.~M.~Wagner,
Nucl.\ Phys.\  {\bf B625}, 345 (2002); 
M.~S.~Carena, J.~R.~Ellis, A.~Pilaftsis and C.~E.~M.~Wagner,
Phys.\ Lett.\  {\bf B495}, 155 (2000).

\end{thebibliography}
\end{document}